\documentclass[a4paper,11pt,floatfix,showpacs,tightenlines,showkeys,superscriptaddress,amsmath,amssymb,nofootinbib]{article}
\pdfoutput=1
\usepackage{jcappub}
\usepackage{amssymb,amsbsy,epsfig,color,graphicx}
\usepackage{color}
\usepackage{[longtable}
\usepackage{array}
\usepackage{dcolumn}   
\usepackage{cellspace}
\usepackage{mathtools}
\usepackage{amstext}
\usepackage{amssymb}
\usepackage{stmaryrd}
\usepackage{stackrel}
\usepackage{graphicx}
\usepackage{esint}
\usepackage[utf8]{inputenc}
\usepackage{blindtext}
\usepackage{float}
\restylefloat{table}
\usepackage{booktabs}
\usepackage{enumitem}

\usepackage{etoolbox} 
\usepackage{lipsum} 
\usepackage[capitalize]{cleveref}

\usepackage{multirow}
\usepackage[caption=false]{subfig}
\renewcommand\[{\begin{equation}}
\renewcommand\]{\end{equation}}

\newcommand{\ba}{\begin{eqnarray}}
\newcommand{\ea}{\end{eqnarray}}

\makeatletter

\appto{\appendix}{%
	\@ifstar{\def\thesection{\unskip}\def\theequation@prefix{A.}}%
	{\def\thesection{\Alph {section}}}%
}
\makeatother

\begin{document}

	\title{Echoes from corpuscular black holes}
	
	\author[]{Luca Buoninfante}
      	\affiliation[]{Department of Physics, Tokyo Institute of Technology, Tokyo 152-8551, Japan}
      	\emailAdd{buoninfante.l.aa@m.titech.ac.jp}

	\date{\today}
	

\abstract{In the corpuscular picture of black hole there exists {\it no} geometric notion of horizon which, instead, only emerges in the semi--classical limit. Therefore, it is very natural to ask -- what happens if we send a signal towards a corpuscular black hole? We show that quantum effects at the horizon scale imply the existence of a surface located at an {\it effective} radius $R=R_s(1+\epsilon)$ slightly larger than the Schwarzschild radius $R_s,$ where $\epsilon=1/N$ and $N$ is the number of gravitons composing the system. Consequently, the reflectivity of the object can be non--zero and, indeed, we find that incoming waves with energies comparable to the Hawking temperature can have a probability of backscattering of order one. Thus, modes can be trapped between the two potential barriers located at the photon sphere and at the surface of a corpuscular black hole, and periodic echoes can be produced. The time delay of echoes turns out to be of the same order of the scrambling time, {\it i.e.}, in units of Planck length it reads $\sqrt{N}\,{\rm log}\,N.$  We also show that the $\epsilon$--parameter, or in other words the compactness, of a corpuscular black hole coincides with the quantum coupling that measures the interaction strength among gravitons, and discuss the physical implications of this remarkable feature.}

\maketitle


\section{Introduction}
Black holes have challenged physicists for very long time and posed very intriguing questions which still lack of definite answers; for instance we can think of the unsolved puzzles of classical singularities~\cite{Hawking} and information loss~\cite{Giddings:2006sj,Unruh:2017uaw,Mathur:2009hf}. 
In the past decades several attempts have been made to find a resolution to these puzzles. In this paper, we discuss a somewhat natural approach in which black holes are treated as composite quantum systems made up of quanta~\cite{Dvali:2011aa,Dvali:2012wq,Dvali:2013eja}. In this picture, black holes of arbitrarily large mass are assumed to be self--sustained gravitational bound states of soft gravitons whose wavelength is of the same order of the size of the system itself. More precisely, a quantum black hole can be represented as a condensate of gravitons stuck at the critical point of a quantum phase transition~\cite{Dvali:2012en}. The quantum criticality is the crucial property in order to explain the microscopic origin of the holographic degrees of freedom~\cite{Dvali:2012en,Dvali:2017nis,Dvali:2018xpy} and to show the logarithmic scaling of the scrambling time~\cite{Dvali:2013vxa} which was previously conjectured in Refs.~\cite{Hayden:2007cs,Sekino:2008he}. One often refers to {\it corpuscular black holes} or {\it black holes' portrait} when adopting such a quantum microscopic description; see also Refs.~\cite{Casadio:2014vja,Casadio:2015bna,Casadio:2017cdv,Casadio:2018qeh,Casadio:2019cux,Buoninfante:2019fwr} for a complementary view and Ref.~\cite{Giusti:2019wdx} for a review.

Let us emphasize that our point of view does not rely on any ultraviolet completion of gravity, therefore it drastically differs in spirit from other approaches based, for example, on string theory like the fuzzball paradigm~\cite{Mathur:2005zp}. Instead, we assume that all the main features of macroscopic black holes can be simply captured in terms of quantum interactions among weakly--interacting soft (low energy) gravitons by using an effective field theoretic approach. In other words, all the whole black hole properties can be explained as originating by a {\it collective} quantum phenomenon.

In this manuscript, we are interested in exploring {\it new} physical aspects of corpuscular black holes, in particular to understand what happens if we send a signal towards them. Quantum mechanically {\it no} geometric notion of horizon can exist, indeed we show that the size of such a gravitational--quantum system does {\it not} strictly coincide with the Schwarzschild radius but is slightly larger due to quantum effects at the horizon scale. This means that the surface of a corpuscular black hole is characterized by a non--zero reflection coefficient. This last property leads to powerful physical implications, like the existence of {\it echoes}~\cite{cardoso-2016b,Cardoso:2016oxy,Abedi:2016hgu}.

The paper is organized as follows. In Section~\ref{corp-BH}, we briefly review the main aspects of the corpuscular picture of a black hole. In Section~\ref{sec-echoes}, we define the effective radius and the compactness of a corpuscular black hole, and show that quantum effects at the horizon scale can be parametrized in terms of $1/N$ corrections, with $N$ being the number of gravitons composing the system. Subsequently, we estimate the reflectivity of a corpuscular black hole and discuss the existence of echoes which are a key property to discriminate between (semi--)classical and quantum black holes.  In Section~\ref{compact-gauge}, we show that the compactness parameter of a corpuscular black hole coincides with the quantum coupling of the graviton--graviton interaction strength, and discuss the physical implications. Section~\ref{conclus-sec} is devoted to summary and concluding remarks.

Throughout the paper we  adopt the mostly positive convention for the metric signature, $(-+++),$ and set $c=1=k_B$ while keeping $\hbar\neq 1\neq G.$ In these units the Planck length and mass are defined as $L_p=\sqrt{\hbar G}$ and $M_p=\sqrt{\hbar/G},$ respectively. Moreover, for simplicity we neglect irrelevant numerical factors of order one and restore them only when needed.

\section{Corpuscular black holes}\label{corp-BH}

The concept of corpuscular black hole is based on the idea of {\it self--completion by classicalization}~\cite{Dvali:2010bf,Dvali:2010jz,Dvali:2016ovn}, according to which a theory can complete itself at some energy scale by producing states of high--multiplicity made up of the same low energy degrees of freedom. All the constituents are soft and still weakly coupled, while their collective behavior can be strong and result in a quasi--classical state due to the very high number of quanta. 

\subsection{Classicalization}

As argued in~\cite{Dvali:2011aa,Dvali:2010bf,Dvali:2011th,Dvali:2014ila}, Einstein's general relativity is an example of theory which can {\it classicalize} and for which the many--particle states would correspond to (corpuscular) black holes made up of a very large number of weakly--interacting soft constituents.

We can illustrate such a phenomenon by adopting a particle physicist point of view and making use of the Hoop Conjecture~\cite{Thorne:1972ji} according to which black holes can be formed in a scattering process between two particles if the impact parameter satisfies $r\lesssim G E_{\scriptscriptstyle CM}$ where $E_{\scriptscriptstyle CM}$ is the centre--of--mass energy. Classicalization states that a corpuscular black hole can be created in a $2\rightarrow N$ scattering process with a final state containing $N\gg 1$ soft weakly--interacting quanta of wavelength $\lambda_g\sim G E_{\scriptscriptstyle CM}$ or, in other words, of energy $m_g= \hbar/\lambda_g\sim M_p^2/E_{\scriptscriptstyle CM}\,.$  Note that, by imposing the conservation of energy, $E_{\scriptscriptstyle CM}=Nm_g,$ we can write $N= E^2_{\scriptscriptstyle CM}/M_p^2.$ 

The gravitational interaction is derivative in nature, therefore we are allowed to define an effective quantum coupling for one single momentum exchange, $\hbar/\lambda_g,$ as follows~\cite{Dvali:2011aa}: 
\begin{equation}
\alpha_g\equiv \frac{\hbar G}{\lambda_g^2}=\frac{L_p^2}{\lambda_g^2}= \frac{M_p^2}{E_{\scriptscriptstyle CM}^2}= \frac{1}{N}\,,\label{single-coupl}
\end{equation}
which represents the self--coupling for each constituent and is always less than one as $N\gg 1.$ Moreover, we can also define a {\rm collective} quantum coupling associated to the whole final $N$--particle state:
\begin{equation}
g\equiv N\alpha_g= 1\,.\label{coll-coupl}
\end{equation}
The last equation is the essence of classicalization: all interacting quanta remain weakly--coupled among each other but their collective behavior can be strong and exhibit non--perturbative features.

So far we have used the word "quanta" without specifying their real nature, indeed they could be either baryons or gravitons, and their number depends on how the final state has been formed. In what follows we assume that all the constituents are pure gravitons since the number of baryons is always negligible and not relevant at least for the purpose of this paper. In relation to this, let us mention that in Ref.~\cite{Casadio:2016zpl} the roles of both matter and gravitons in a gravitational collapse were discussed, and it is was shown that the numbers of gravitons $N$ is always a lot larger than the number of baryons $N_B,$ indeed one has $N\sim N_B^2(m_B/M_p^2)^2\gg N_B,$ with $m_B<M_p$ being the mass of a single baryon. For instance, for a solar mass compact object, $M\sim 10^{38}M_p,$ we would have $N_B\sim 10^{57}$ and $N\sim 10^{76}.$

\subsection{Self--sustained gravitational quantum state}\label{sec-self-sust}

Let us now show that it is indeed possible to form a gravitationally bound state of many gravitons whose number can be arbitrarily large. 
Given a system of $N$ weakly--interacting soft gravitons of wavelength $\lambda_g= \hbar/m_g\gg L_p,$ the total mass is
\begin{equation}
M=Nm_g, 
\label{tot-mass}
\end{equation}
and each graviton interacts gravitationally with the others $N-1\simeq N$ through the potential $\Phi(r)= -GNm_g/r\,.$ Moreover, from the classicalization argument we have learnt that the wavelength of each single constituent is of the same order of system's size, therefore the binding energy felt by each graviton reads $U_g= m_g\Phi(\lambda_g)= -N\hbar\alpha_g/\lambda_g,$ while their kinetic energy can be approximated by $E_g\simeq m_g\,.$ By imposing the energy balance equation $E_g+U_g= 0,$ we can find the necessary and sufficient condition to form a self--sustained gravitationally bound system \cite{Dvali:2011th,Casadio:2014vja}:
\begin{equation}
m_g-\frac{N\hbar\alpha_g}{\lambda_g}= 0\quad  \Leftrightarrow \quad N\alpha_g= 1\,,\label{balance}
\end{equation}
which coincides with the relation~\eqref{coll-coupl}. By using Eq.~\eqref{single-coupl} we can write the total mass of the system and the Schwarzschild radius as
\begin{equation}
M=\sqrt{N}M_p,\quad R_s= 2\sqrt{N}L_p\,.\label{mass-and-sch}
\end{equation}
From Eqs.~\eqref{balance} and~\eqref{mass-and-sch}, it follows that the $N$ weakly--interacting gravitons $(\alpha_g= 1/N\ll 1)$ form a condensate of attractive bosons, indeed they are soft and their wavelength is of the same order of the size of the system $(\lambda_g\sim R_s)\,.$ Moreover, their collective coupling $g=N\alpha_g$ is always equal to one meaning that such a condensate is stuck at a critical point of a phase transition~\cite{Dvali:2012en}.

In the black hole corpuscular picture, the Hawking radiation can be explained as a leakage phenomenon exhibited by the system which is forced to lose quanta in order to remain stuck at the critical point and balance the collapse of the condensate~\cite{Dvali:2012en}. Indeed, each graviton has a non--zero probability of escaping from the bound state due to the scattering with the remaining $N-1,$ and it is easy to understand that the energy and wavelength thresholds are~\cite{Dvali:2011aa} 
\begin{equation}
E_{\rm esc}=\frac{\hbar}{\sqrt{N}L_p}\,,\quad \lambda_{\rm esc}= \sqrt{N}L_p\,. \label{escape-scales}
\end{equation}
The depletion rate $\Gamma$ is approximatively given by a product involving the coupling squared $\alpha_g^2$, the characteristic energy $E_{\rm esc}$ and the combinatoric factor $N(N-1)\simeq N^2$ (for $N\gg 1$)~\cite{Dvali:2011aa}, and after some simple algebra one can show that
\begin{equation}
\Gamma= \displaystyle \frac{\hbar}{\sqrt{N}L_p}+\mathcal{O}\left(\frac{\hbar}{L_p}\frac{1}{N^{3/2}}\right)\,, \label{depl-rate}
\end{equation}
where the second term is a higher order contribution to $2\rightarrow2$ scattering containing three vertexes, it is proportional to
$\alpha_g^3N^2E_{\rm esc}$ and represents a pure quantum correction which is absent semi--classically. 

By using Eq.~\eqref{depl-rate}, the evaporation rate of a corpuscular black hole can be expressed as a function of $N$ and understood as a loss of gravitons through the depletion process:
\begin{equation}
\dot{N}= - \frac{1}{\sqrt{N}L_p}+\mathcal{O}\left(\frac{1}{L_p}\frac{1}{N^{3/2}}\right)\,, \label{evaporation rate}
\end{equation}
where the dot stands for derivative with respect to the Schwarzschild time $t$. The higher order corrections in Eqs.~\eqref{depl-rate} and~\eqref{evaporation rate} play a crucial role for the resolution of the information loss paradox as they can be interpreted as non--thermal corrections which allow to recover consistently the information in a finite amount of time~\cite{Dvali:2012wq,Dvali:2015aja}. By making a comparison with the Stefan--Boltzmann law of a semi--classical black hole, we can also define the temperature for the condensate as a function of the number of gravitons $N:$
\begin{equation}
T_H=\frac{\hbar}{\sqrt{N}L_p}\,, \label{temperature}
\end{equation}
which coincides with the Hawking temperature once we explicit the mass through $\sqrt{N}=M/M_p.$

The presence of a quantum critical point naturally implies the emergence of $N$ gapless modes, as it happens for any condensate of attractive bosons in proximity of a phase transition~\cite{Bogolyubov:1947zz}. In Refs.~\cite{Dvali:2012en,Dvali:2017nis,Dvali:2018xpy}, by working with a scalar toy--model it was argued that the $N$ gapless modes can play the role of holographic degrees of freedom which carry the information stored in the system, and give a microscopic counting of the black hole entropy:
\begin{equation}
S= N.\label{entropy}
\end{equation}

\section{Existence of echoes} \label{sec-echoes}

The main question we ask in this Section is -- {\it what happens if we send a signal towards a corpuscular black hole?} To find an answer we need to understand whether quantum mechanically the surface of a corpuscular black hole is characterized by some non--vanishing reflection coefficient.

\subsection{Effective radius} 

In the semi--classical scenario, everything we throw inside the black hole will never come out again because of the horizon. However, in the scenario depicted above it so happens that quantum mechanically there exists {\it no} geometric notion of horizon, for instance soft quanta can exceed the threshold energy and escape from the bound state. This suggests that there is {\it no} physical boundary which disconnects the inside and outside of a corpuscular black hole. Therefore, we would also expect that an ingoing particle could have a non--zero probability of backscattering after having interacted with a corpuscular black hole.
Hence, the size of the graviton condensate can not be strictly identified with the Schwarzschild radius $R_s=\sqrt{N}L_p$ as there are always quantum fluctuations which make the horizon a {\it fuzzy} region.  

In fact, the authors in Refs.~\cite{Casadio:2014vja,Casadio:2015bna}, by using tools of horizon quantum mechanics~\cite{Casadio:2013tma,Casadio:2016fev} through which a quantum wave--function is associated to the location of system's surface, it was explicitly shown that the Schwarzschild radius is corrected by terms proportional to $R_s/\sqrt{N},$ and analogously that the energy deviates from the black hole mass by a Hawking mode energy $M_p/\sqrt{N}.$ Obviously, in the exact semi--classical limit $N\rightarrow \infty,$ the deviations vanish and the geometric notion of horizon is recovered.

Such a result can be naively understood as follows. Quantum mechanically black holes possess a non--zero temperature~\eqref{temperature} and, because of the quantum back--reaction of the Hawking radiation on the black hole, a distant observer at infinity would see a hot surface of energy $E=M+\delta E,$ with $\delta E\sim T_H\sim M_p/\sqrt{N}$ being the thermal fluctuation which vanishes in the (semi--)classical limit. As a consequence, the size of the system would be given by an {\it effective} radius  $R=2GE=R_s+\delta R,$ with $\delta R\sim L_p/\sqrt{N},$ which can be also recast as 
\begin{equation}
R=R_s\left(1+\epsilon\right)\,,\quad {\rm with}\quad \epsilon\equiv\frac{1}{N}\ll1\,, \label{effect-radius}
\end{equation}
where the $1/N$ corrections are intrinsically quantum in nature and drastically distinguish the quantum corpuscular picture from the (semi--)classical one.

Let us emphasize that the correction $\delta R$ was defined in Schwarzschild coordinates, but can be also expressed in proper distance. By making a geometric computation, which is well justified as long as we are outside the condensate, we can straightforwardly check that the fluctuation $\delta R=R_s/N$ corresponds to a proper Planck distance:
\begin{equation}
\int_{R_s}^{R_s+\delta R} \frac{{\rm d}r}{\sqrt{f(r)}}\sim \sqrt{R_s\delta R}\sim L_p\,,\label{proper distance}
\end{equation}
where 
\begin{equation}
f(r)=1-\frac{R_s}{r}\label{metric-compon}
\end{equation}
and we have used $N\gg 1$ ({\it i.e.}, $\epsilon\ll 1$). 

Note that we can obtain the same result by reminding that $T_H$ is the Hawking temperature as measured by an observer at infinity, while the local temperature on the surface of a corpuscular black hole reads
\begin{equation}
T_R=\frac{T_H}{\sqrt{f(R)}}\sim \frac{1}{2G}\sqrt{R_s\delta R}\sim \frac{L_p}{2G}\,, \label{local temperature}
\end{equation}
which means that locally the quantum back--reaction causes Planck scale corrections, $R= R_s+2GT_R\sim R_s+L_p.$ In other words, Hawking quanta of wavelength $L_p$ (or mass $M_p$) are produced in the vicinity of the surface and, indeed, define the thickness of the hot membrane above the would be horizon.

What we have just discussed is consistent with the fact that quantum gravity corrections are expected to become important within a Planck length from the horizon as predicted by the stretched horizon~\cite{Susskind:1993if}, "brick" wall~\cite{tHooft:1984kcu} and fuzzball~\cite{Mathur:2005zp} models which share the common feature of a microscopic structure in the near horizon region.

Note that the radius in Eq.~\eqref{effect-radius} can be also written as only a function of the number of gravitons $N,$ namely
\begin{equation}
R=2L_p\left(\sqrt{N}+\frac{1}{\sqrt{N}}\right)\,. \label{radius-only-N}
\end{equation}
This last expression is very intriguing as it explicitly shows that the radius is invariant under the transformation $N\leftrightarrow 1/N,$ which connects large and small scales. Indeed, by writing $R=R_s+\lambda_s,$ with $\lambda_s=\hbar/M$ being the De Broglie wavelength of the entire system, the transformation reads $R_s\leftrightarrow \lambda_s.$  This peculiar symmetry might have some important consequence and surely deserve future investigations.

\subsection{Compactness}

We now introduce the {\it compactness} parameter
\begin{equation}
\mu\equiv 1-\frac{R_s}{R}=\frac{\epsilon}{1+\epsilon}\,; \label{compactness}
\end{equation}
the smaller $\mu$  is, the more compact the system is. Note that, the (semi--)classical black hole $(\epsilon =0)$ is the most compact object that can exist in nature and is characterized by $\mu_{\rm BH}=0,$\footnote{Classically, the presence of a singularity suggests that the quantity $R_s/R$ can also become larger than one, {\it i.e.}, $\mu$ can be negative. Indeed, if we define $R$ as the minimal radius containing some matter distribution, it so happens that after crossing the horizon all matter collapses into the singularity so that $R_s/R\rightarrow \infty.$ However, our point view relies on the fact that physics at distances $R<R_s$ can not be probed, thus $R_s$ is the minimal radius defining a physical boundary beyond which no measurement can be made. In this sense, we assume that $R_s/R=1$ $(\mu=0)$ corresponds to the maximal compactness scenario, which is the case for (semi--)classical black holes. See Ref.~\cite{Casadio:2020ueb} where quantum mechanical arguments to prevent $R_s/R>1$ $(\mu<0)$ were put forward.} while in presence of quantum corrections we have
\begin{equation}
\mu\simeq \epsilon=\frac{1}{N}\,, \label{small-eps-compact}
\end{equation}
where to go from~\eqref{compactness} to~\eqref{small-eps-compact} we have used $\epsilon=1/N\ll 1.$ Therefore, the compactness parameter of a corpuscular black hole assumes a very simple form and only depends on the number of gravitons composing the gravitational bound state.

The absence of a horizon also means that, in principle, the surface of a corpuscular black hole is {\it not} a completely absorbing membrane but can possess a non--vanishing reflection coefficient.

\subsection{Reflectivity}

Let us consider an observer who sends an ingoing wave which can reach the surface of our astrophysical object. For a (semi--)classical black hole, the wave never comes back after crossing the horizon. However, if the central object is horizonless, it may happen that a fraction of the wave interacts with its surface, gains a non--zero probability of being reflected and escapes to spatial infinity. Such a process can be described in terms of a Schr\"odinger--like equation
\begin{equation}
\frac{{\rm d}^2\psi(x)}{{\rm d}x^2}+[\omega^2-V(r)]\psi(x)=0\,, \label{wave-eq}
\end{equation}
where $\omega$ is the frequency of the wave $\psi$ and $x=r+2GM{\rm log}(r/2GM-1)$ is the tortoise coordinate. We do not write down the explicit expression of $V(r(x))$ which corresponds to the well known Regge--Wheeler (axial)~\cite{Regge:1957td} or Zerilli (polar) potential~\cite{Zerilli:1970se} for spin--$2$ perturbations.  Very close to the surface (and for $\epsilon\ll 1$) the solution $\psi(x)$ can be expressed as a combination of ingoing and outgoing waves,
\begin{equation}
\psi(x)=A_{in}e^{-i\omega x}+A_{out}e^{i\omega x}\,, \label{wave-combination}
\end{equation}
where $A_{in}$ and $A_{out}$ are the ingoing and outgoing amplitudes, respectively, and typically are complex and frequency--dependent. We can define the reflection probability
\begin{equation}
|\mathcal{R}(\omega)|^2=\left|\frac{A_{out}}{A_{in}}\right|^2\,, \label{reflectivity}
\end{equation}
which is also known as {\it reflectivity,} and it is equal to zero in case of total absorption $(A_{out}=0)$ and to one in case of perfect reflection. Analogously, we can define the absorption probability as $|\mathcal{T}|^2=1-|\mathcal{R}|^2.$ More generally, the reflectivity of an object includes both {\it elastic} and {\it inelastic} contributions: the former refers to backscattering happening in the outside region, for instance after hitting the surface; while the latter to waves which can pass through the surface, interact with the interior and come out again~\cite{Carballo-Rubio:2018jzw}. 

We can estimate the reflectivity in Eq.~\eqref{reflectivity} by noticing that it must coincide with the probability of backscattering~\cite{Guo:2017jmi} which, in turn, is proportional to the product of interaction probability, $P_{int},$ and escape probability, $P_{esc}:$
\begin{equation}
|\mathcal{R}|^2= P_{int}\cdot P_{esc}\,.\label{tot-prob}
\end{equation}

In our framework the relevant degrees of freedom are soft gravitons which populate both the interior region of a corpuscular black hole and its atmosphere (Hawking radiation) up to the surface whose structure is made up of quanta of local wavelength $L_p.$ Thus, since we are mainly interested in gravitational perturbations to the system, the relevant process that can take place is given by scattering of an incoming graviton with the gravitons living either inside or outside the system.

Let us compute the reflectivity for both elastic and inelastic cases, separately.
\begin{itemize}
	
	\item In the elastic case, $P_{int}(\rho)$ corresponds to the probability of an infalling graviton to interact with Hawking quanta during its trip from some proper distance $D\gg \rho,$ that we can assume to be $D=\infty$ (without any loss of generality), up to a proper distance $\rho\geq L_p$ from the would be horizon. The saturation $\rho=L_p$ corresponds to the time at which the incoming graviton hits the surface. More specifically, the interaction probability is defined as
	\begin{equation}
	\qquad\, P_{int}(\rho)=\int_\rho^\infty{\rm d}\rho'\, \frac{{\rm d} P_{int}(\rho')}{{\rm d}\rho'}\sim \int_\rho^\infty{\rm d}\rho'\, \sigma(\rho') n(\rho')\,, \label{Pint}
	\end{equation}
	where $\sigma(\rho')$ is the cross--section and $n(\rho')$ the density of Hawking quanta at the proper location $\rho'$. At the leading order it is sufficient to consider $2\rightarrow 2$ scattering processes and, for simplicity, we also assume that the final energy of each scattered particle is equal to its initial value. 
	
	By working in the centre--of--mass frame, the cross--section reads~\cite{Guo:2017jmi}
	\begin{equation}
	\sigma (\rho)\sim G^2E^2_{\scriptscriptstyle CM}\sim \frac{L_p^4}{\rho^2}\frac{\hbar \omega}{T_H}\,,
	\end{equation}
	where $E_{\scriptscriptstyle CM}(\rho)\sim\sqrt{T(\rho)\hbar \omega(\rho)}$ and we have used the fact that the local energies of Hawking quanta and incident graviton are $T(\rho)\sim \hbar/\rho$ and $\hbar\omega(\rho)\sim R_s\hbar\omega/\rho,$ respectively.
	
	The number density of gravitons in the radiation is
	\begin{equation}
	n(\rho)\sim \frac{T(\rho)^3}{\hbar^3}\sim \frac{1}{\rho^3}\,,
	\end{equation}
	so that after integration we get the following expression for the interaction probability~\eqref{Pint}:
	\begin{equation}
	P_{int}\sim \frac{L_p^4}{\rho^4}\frac{\hbar \omega}{T_H}\,.
	\end{equation}

	Furthermore, the probability $P_{esc}$ is proportional to the solid angle $\Delta \Omega(\rho)$ under which particles of wavelength $\lambda\ll R_s$ (geometric optic approximation) can escape from the compact object~\cite{Cardoso:2017njb,Guo:2017jmi,Carballo-Rubio:2018jzw}
	\begin{equation}
	P_{esc}\sim \Delta \Omega(\rho)\sim \left(\frac{\rho}{R_s}\right)^2\,.\label{solid-ang}
	\end{equation}
	Note that in the opposite regime $(\lambda \gtrsim R_s)$ a wave would lay on a region larger than the central object's size, so that the previous formula would lose physical meaning. In fact, most likely such long wavelengths waves would be able to escape after having interacted, {\it i.e.}, we can write $P_{esc}\sim 1.$
	
	We now estimate the total elastic contribution to the reflectivity in two different regimes. For incoming energies $\hbar \omega\gg T_H$ (high frequency) Eq.~\eqref{solid-ang} is valid and we obtain
	\begin{equation}
	|\mathcal{R}|^2\sim \frac{L_p^4}{R_s^2\rho^2}\frac{\hbar \omega}{T_H}\sim \frac{L_p^4}{\rho^2\lambda R_s}\,,
	\end{equation}
	which can be of order one only when 
	\begin{equation}
	\rho^2\sim \frac{L_p^4}{\lambda R_s}\,,\label{eq-TH<ho}
	\end{equation}
	with $\lambda=1/\omega$  being the incoming graviton wavelength. Since $\lambda>L_p$ ({\it i.e.}, $\hbar \omega<M_p$) and $R_s\gg L_p,$ Eq.~\eqref{eq-TH<ho} gives $\rho <L_p$  which, however, can not happen as by construction $\rho\geq L_p.$ Therefore, we have learnt that for incoming energies larger than Hawking temperature the probability of backscattering is very low,
	\begin{equation}
	|\mathcal{R}|^2\lesssim \frac{1}{\sqrt{N}}\frac{L_p}{\lambda}=\frac{1}{\sqrt{N}}\frac{\hbar \omega}{M_p}\ll 1\,.
	\end{equation}

	Whereas for energies $\hbar\omega\lesssim T_H$ (low frequency) or, in other words for wavelengths  $\lambda\gtrsim R_s,$ the reflectivity reads $(P_{esc}\sim 1)$
	\begin{equation}
	|\mathcal{R}|^2\sim \frac{L_p^4}{\rho^4}\frac{\hbar \omega}{T_H}\,,
	\end{equation}
	which can be of order one, $|\mathcal{R}|^2\sim 1,$ when $\rho \simeq L_p$ and $\hbar \omega \simeq T_H.$ This means that an incoming graviton with energy comparable to the Hawking temperature and that hits the surface of a corpuscular black hole can have a probability of backscattering of order one. It so happens that gravitational waves emitted from binary merger have wavelengths of the same order of the Schwarzschild radius, {\it i.e.}, $\omega\sim 1/R_s\sim T_H/\hbar,$ which coincides with what we would need in order to have a sufficiently large reflectivity. This is a remarkable result and opens a new window of opportunity to test the corpuscular picture with gravitational wave experiments, as we will discuss below. It is also worthwhile emphasizing that to some extent our result agrees with the approach in Ref.~\cite{Oshita:2019sat}.
	
	\item We can now ask what happens to those gravitons which failed to backscatter in the outside region, namely whether they still have some chance to backscatter after having interacted with the interior (inelastic reflection). We expect such a probability to be quite small because of  the very large number of states $(\sim e^N)$ that such an infalling quanta should explore before coming out again.

	Inside a corpuscular black hole the characteristic length and energy scales are given by $R_s$ and $\hbar /R_s\sim T_H,$ so that we can estimate $\sigma\sim L_p^2\alpha_g,$ $n\sim N/R_s^3$ and $P_{esc}\sim 1,$ which give 
	\begin{equation}
	|\mathcal{R}|^2\sim \frac{L_p^2}{R_s^2}\sim \frac{1}{N}\ll 1\,.
	\end{equation}
	We can also make a rough estimation of the rate of interaction  between the infalling graviton and the $N$ gravitons composing the system: $\Gamma_{int} \sim N\alpha_g^2 \hbar/R_s\sim M_p/N^{3/2},$ where $\alpha_g^2$ is the interaction strength, $\hbar/R_s$ the characteristic energy scale and $N$ a combinatoric factor. This means that only after a time $\Delta t_{int}\sim L_p N^{3/2}\sim L_p M^3/(M_p^3)$ the interaction process would become significant. However, such a time scale is of the same order of the black hole life time, therefore for all practical purposes the inelastic reflectivity turns out to be zero.
	
\end{itemize}

\vspace{0.18cm}

To summarize, we have shown that the reflectivity of a corpuscular black hole can be of order one only for infalling gravitons with energies $\hbar\omega\simeq T_H$ and that can travel up to $\rho=L_p$. This means that a low frequency graviton can backscatter only after having hit the hot membrane of a corpuscular black hole located at $R=R_s+L_p$ (in proper distance). This is also consistent with the definition of reflection coefficient, $\mathcal{R}=A_{out}/A_{in},$ based on the decomposition in Eq.~\eqref{wave-combination} which is only valid close to the surface ($\rho\simeq L_p$). In all other cases the reflectivity turns out to be negligible and incoming gravitons get absorbed. Note that in the (semi--)classical limit the above discussion would not hold because in presence of a horizon {\it no} real surface would exist and, consistently with the equivalence principle, an infalling particle would feel no interaction and simply get absorbed.

Let us emphasize that through our analysis we were only able to make an estimation based on sensible physical arguments, and thus we have obtained an approximate expression of the reflectivity for an incoming spin--$2$ wave. However, in order to confirm and strengthen our conclusions a more rigorous study is surely needed, to include also perturbations of different spin. For instance, in Ref.~\cite{Oshita:2019sat} it was argued, in a different context, that any low frequency $\hbar\omega\lesssim T_H$ would be characterized by a large reflectivity. However, we expect that waves with energy lower than Hawking temperature will have a very low probability to reach spatial infinite because of the potential barrier at the photon sphere whose height is proportional to $T_H^2$. Hence, the probability of backscattering and reaching spatial infinite can be of order one only for waves of energies $\hbar \omega\simeq T_H.$

\subsection{Echoes}\label{echoes-corp}

It is now clear that quantum mechanically a black hole has to be thought of as a {\it horizonless} gravitational bound system whose surface is characterized by a non--zero reflection coefficient. Indeed, unlike high frequency modes $(\hbar \omega\gg T_H)$ which would excite the degrees of freedom on the surface and get absorbed, it so happens that lower frequencies $(\hbar \omega\simeq T_H)$ can have a high probability of backscattering. The main physical implication of this feature is the production of {\it echoes}~\cite{cardoso-2016b,Cardoso:2016oxy,Abedi:2016hgu} whose amplitude is roughly proportional to the reflectivity. In the absence of a horizon, indeed, waves can be trapped between the two potential barriers located at the surface and at the photon sphere, and periodically come out with decreasing amplitude in time. See also Refs.~\cite{Holdom:2016nek,Abedi:2018npz,Cardoso:2017cqb,Cardoso:2017njb,Carballo-Rubio:2018jzw,Guo:2017jmi,Maggio:2017ivp,Maggio:2018ivz,Testa:2018bzd,Maggio:2019zyv,Buoninfante:2019swn,Buoninfante:2019teo,Cardoso:2019rvt,Foit:2016uxn,Cardoso:2019apo,Coates:2019bun,Oshita:2019sat,Wang:2019rcf,Buoninfante:2020cqz,Abedi:2020ujo,Salvio:2019llz,Konoplya:2018yrp,Bronnikov:2019sbx,Wang:2018mlp,Wang:2019szm,Maggio:2020jml} and references therein for recent theoretical and phenomenological studies on echoes from black hole mimickers.

\begin{figure}[t]
	\includegraphics[scale=0.5]{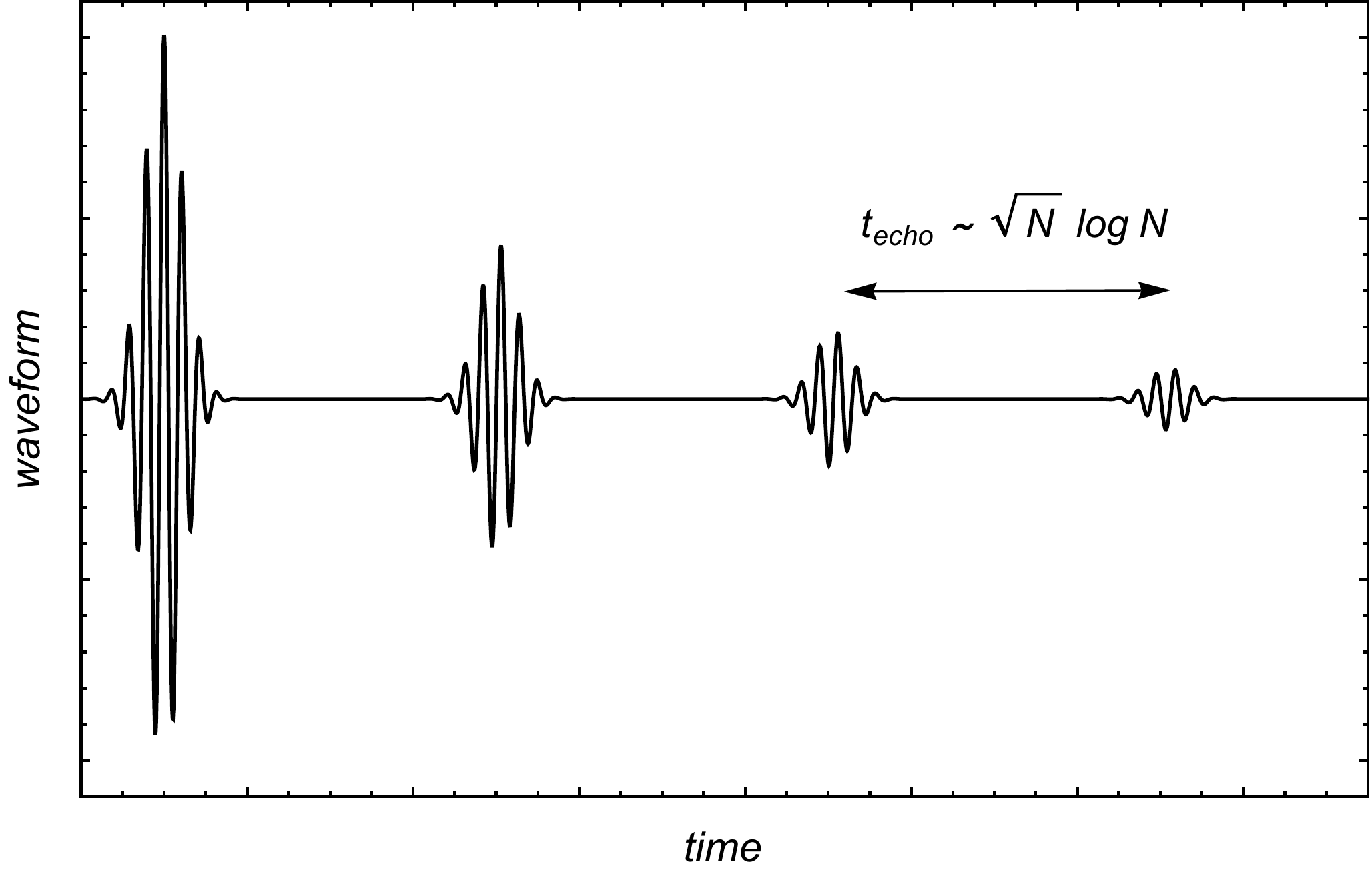}
	\centering
	\protect\caption{Illustration of an echo signal; $t_{echo}\sim \sqrt{N}\, {\rm log}\,N$ is the period in units of Planck length, where $N$ is the number of gravitons.}\label{fig1}
\end{figure}

A typical physical configuration in which such a phenomenon becomes relevant is during a binary merger. If the resulting object is devoid of any horizon (for instance a corpuscular black hole), the corresponding waveform will be characterized by additional periodic pulses after the prompt ringdown~\cite{cardoso-2016b,Cardoso:2016oxy,Abedi:2016hgu}, thus representing a smoking gun signature of quantum effects at the horizon scale; see Fig.~\ref{fig1} for an illustration.

A crucial physical quantity is the period $t_{echo}$ that corresponds to the time for the roundtrip from the photon sphere to the surface; in the $\epsilon\ll 1$ limit it reads~\cite{cardoso-2016b,Cardoso:2016oxy,Abedi:2016hgu}:
\begin{equation}
t_{echo}=\displaystyle 2\int_{R_s(1+\epsilon)}^{\sim 3GM} \frac{{\rm d}r}{f(r)}\sim -2R_s {\rm log} \,\epsilon\,,
\label{echo-time}
\end{equation}
where $f(r)$ is given in Eq.~\eqref{metric-compon}. The {\it echo time} for a corpuscular black hole can be expressed in terms of the number of gravitons $N,$ and by reinserting the correct numerical factors we get
\begin{equation}
t_{echo}= 4L_p\sqrt{N}\,{\rm log}N\,,
\end{equation}
The above functional form suggests that the number of gravitons $N$ controls the {\it rhythm} at which a quantum black hole relaxes after being perturbed; see Fig.~\ref{fig1}.

Very interestingly, the echo time turns out to be of the same order of the scrambling time which is given by $t_{scr}=(\hbar \beta/2\pi){\rm log} N$ where $\beta=1/T,$ and that corresponds to the time needed for the information contained in a perturbation to spread among the $N$ degrees of freedom and get lost in the system~\cite{Hayden:2007cs,Sekino:2008he}. Indeed, in Ref.~\cite{Dvali:2013vxa} it was explicitly shown that the scrambling time of such a graviton condensate scales logarithmically with $N,$ {\it i.e.}, $t_{scr}\sim R_s{\rm log}\,N,$ and since $\hbar \beta \sim R_s \sim  \sqrt{N}L_p,$ it follows
\begin{equation}
t_{echo}\sim \,t_{scr}\,.
\end{equation}
It is worthwhile mentioning that such a common feature between echoes and scrambling in the context of Planckian corrections at the horizon scale was pointed out for the first time in~\cite{Abedi:2016hgu} and further investigated in~\cite{Saraswat:2019npa}.

Before concluding this Section, let us mention that strong claims have been made on the detection of echoes in some gravitational wave event, but the debate is still open and of course more work is still needed in order to reach a final conclusion~\cite{Abedi:2016hgu,Conklin:2017lwb,Abedi:2018npz,Ashton:2016xff,Westerweck:2017hus}. However, we would like to emphasize that gravitational wave astronomy could offer a new window of opportunity to test models of horizonless compact objects, and in particular the black hole corpuscular picture.

\section{{\large $\epsilon$}--parameter as a quantum coupling}\label{compact-gauge}

In this Section we aim to discuss a novel feature which can uniquely distinguish corpuscular black holes from other kind of horizonless compact objects already existing in the literature. The key physical quantity of our above analysis was $\epsilon$ introduced in Eq.~\eqref{effect-radius}, which coincides with the compactness parameter as we are working in the limit $\epsilon\ll 1,$ see Eq.~\eqref{small-eps-compact}.

As mentioned in Ref.~\cite{Cardoso:2019rvt}, one of the most outstanding problems in the context of horizonless compact objects is the fact that $\epsilon$ is often a fixed parameter and not derived from first principles. Classical black holes are very special because the mass $M$ is a free integration constant, therefore the Schwarzschild metric can describe black holes of any arbitrarily large mass. However, in the case of horizonless spacetimes it often happens that to describe some specific matter configuration or physics beyond Einstein's general relativity one is forced to introduce a new physical scale on which the parameter $\epsilon$ will depend. As a consequence the range of possible values over which the mass can run is limited, so that there exists a critical mass value above which the horizon can not be avoided\footnote{Namely, if $g_{00}=-(1+2\Phi)$ and $g_{11}=(1+2\Psi)^{-1}$ are the two metric components, it can happen that for some critical value of the mass $2\Phi(r^*)=-1$ and $2\Psi(r^*)=-1$ which imply the existence of a horizon at $r^*$.}. The only way one can resolve such an issue is to have a mass--dependent compactness parameter so that by increasing $M$ also $\epsilon$ will change in such a way to preserve the "no horizon" condition.

Most of the approaches lack of a supporting consistent quantum theory and rely on approximate geometric descriptions in which the Schwarzschild metric is modified for radii $r\leq R,$ where $R$ is the size of the object. However, we want to stress that there is no reason why quantum effects should simply act as modifications of the spacetime metric, indeed we believe that such a way of thinking does not capture the real quantum nature of the gravitational interaction. In fact, in order to make a fully quantum treatment we should assume quantum effects to take place on the top of the geometric background.

The corpuscular picture can successfully fit in the above discussion, indeed the compactness parameter of a corpuscular black hole is characterized by the following unique functional dependence: we have $\epsilon=1/N=M_p^2/M^2,$ and the "no horizon" condition can be preserved for any finite $M$ (or, equivalently, for any finite $N$). Let us also emphasize that the specific dependence on the mass squared inverse is the only possibility that would correspond to a proper Planck distance from the would be horizon. Any other inverse power $1/M^n$ with $n\neq 2$ would give different results for the proper distance in Eq.~\eqref{proper distance} which, instead, would depend on the mass itself and not be simply a constant proportional to $L_p$. 

The corpuscular picture is even richer, indeed by using the relation $\alpha_g=1/N,$ the quantity $\epsilon$ simply becomes
\begin{equation}
\epsilon=\alpha_g\equiv {\rm quantum\,\,coupling}\,,\label{compact-coupling}
\end{equation}
which tells us that the $\epsilon$--parameter, or in other words the compactness, {\it coincides} with the quantum gravitational coupling $\alpha_g(R_s)= 1/N$ evaluated at the energy scale $\hbar/R_s,$ see Eq.~\eqref{single-coupl}. It physically means that the compactness of a quantum black hole is {\it solely} controlled by the interaction strength between gravitons: the stronger the interaction is, the less compact a corpuscular black hole will be. It goes to zero only in the semi--classical regime where the intrinsic corpuscular quantum nature is obviously absent. This is a remarkable result and could be a very general property of quantum black holes, that is physically the $\epsilon$--parameter is in one--to--one correspondence with the quantum interaction strength of its microscopic degrees of freedom.

In this respect, the $\epsilon\rightarrow 0$ limit, or in other words the $1/N\ll 1$ corrections, could be understood analogously to the 't Hooft limit in QCD~\cite{tHooft:1973alw} and, this peculiar aspect will surely deserve further investigations; see also Ref.~\cite{Dvali:2020wqi} for an explanation of the close analogy between the semi--classical limit of black holes and the 't Hooft limit. Recently, an effective field theory approach was developed to study horinzonless astrophysical objects and echoes~\cite{Burgess:2018pmm}. A similar framework might turn out to be very suitable in order to treat quantum effects at horizon scale and show that the $\epsilon$--parameter is energy (mass) dependent and can run with it. We leave a concrete realization of such an idea for future works.

\section{Concluding remarks \& outlook}\label{conclus-sec}

In this paper we explored novel aspects of the corpuscular quantum picture of a black hole by understanding what happens to a signal that is sent towards a corpuscular black hole. Quantum mechanically there exists no geometric notion of horizon and because of quantum back--reaction the size of the system is given by an effective radius which turns out to be slightly larger than the usual Schwarzschild radius. This feature drastically discriminates between (semi--)classical and quantum black holes, and has very powerful implications.

In the absence of a horizon, the surface of a corpuscular black hole turns out to be characterized by a non--zero reflection coefficient which remarkably can be of order one for infalling gravitons with energies comparable to the Hawking temperature. Whereas in all other cases the reflectivity turns out to be equal to zero for all practical purposes as it is proportional to negative powers of the entropy.
The main physical consequence of this property is the existence of periodic echoes whose time scale is of the same order of the scrambling time of the system, and that can be expressed simply in terms of the number of gravitons, {\it i.e.}, (in units of Planck length) $t_{echo}\sim \sqrt{N}{\rm log}\,N.$

Current experiments on binary mergers and gravitational waves are seeking for this kind of effects beyond Einstein's general relativity; in fact, the time scale of the delay between echoes is now an accessible measurable quantity to LIGO/LISA~\cite{Cardoso:2019rvt,Abedi:2020ujo}. Therefore, there are stimulating hopes that the black hole corpuscular picture can be really tested in future gravitational wave experiments. 

We also discussed on a remarkable property of corpuscular black holes which distinctly distinguish them from other existing models of horizonless compact objects. We noticed that the $\epsilon$--parameter coincides with the quantum coupling $\alpha_g,$ suggesting a very intriguing connection between the quantum structure of a black hole and its compactness. When $\alpha_g=0$  $(N\rightarrow \infty)$ the quantum interaction among its microscopic degrees of freedom vanish and we consistently recover the (semi--)classical black hole limit in which $\epsilon=0.$ Therefore, the presence of such a quantum nature is what controls the compactness of a corpuscular black hole, offering a natural mechanism to avoid the horizon only based on first principles of general relativity and quantum mechanics combined together.

Let us emphasize that we only worked in a static scenario, therefore further investigations are surely needed in order to understand whether and how a non--zero angular momentum would change our results and predictions.

Before concluding let us mention that an interesting open question that we would like to address and possibly answer in the near future is the following -- can we build models of analogue gravity to verify the existence of echoes and other quantum effects at the horizon scale through experiments in a laboratory? 
Some interesting works have been already done along this direction; indeed, in Ref.~\cite{Torres:2018dso} the existence of quasi-normal modes was experimentally verified  in an analogue gravity experiment by working with classical hydrodynamical systems and simulating the unstable light-ring on the photon sphere. In our case we should deal with quantum condensates, and we expect that by taking into account quantum effects (like depletion) echoes can be simulated. However, a concrete realization of such a physical setup still needs to be worked out and it will certainly be the subject of future investigations.

\acknowledgments
It is a pleasure to thank Paolo~Pani for a critical reading of this manuscript and for very useful and stimulating comments, and Francesco~Di~Filippo for discussions. I am grateful to Vitor~Cardoso, Roberto~Casadio and Andrea~Giusti for suggestions. I would also like to thank the Cosmology Group at Titech for offering a perfect environment where to freely think and enjoy physics. 
This work is supported by JSPS and KAKENHI Grant-in-Aid for Scientific Research No.~JP19F19324.


\begin{thebibliography}{0}
	
	
	\bibitem{Hawking}
	S.~W.~Hawking and G.~F.~R.~Ellis,
	\emph{The Large Scale Structure of Space-Time}, 
	Cambridge University Press, Cambridge (2011).
	
	\bibitem{Giddings:2006sj}
	S.~B.~Giddings,
	Phys.\ Rev.\ D {\bf 74}, 106005 (2006).
	
	\bibitem{Unruh:2017uaw}
	W.~G.~Unruh and R.~M.~Wald,
	Rept.\ Prog.\ Phys.\  {\bf 80}, 092002 (2017).
	
	
	\bibitem{Mathur:2009hf} 
	S.~D.~Mathur,
	Class.\ Quant.\ Grav.\  {\bf 26}, 224001 (2009).
	
	
	
	\bibitem{Dvali:2011aa} 
	G.~Dvali and C.~Gomez,
	Fortsch.\ Phys.\  {\bf 61}, 742 (2013). 
	
	\bibitem{Dvali:2012wq} 
	G.~Dvali and C.~Gomez,
	arXiv:1212.0765 [hep-th].
	
	\bibitem{Dvali:2013eja} 
	G.~Dvali and C.~Gomez,
	JCAP {\bf 1401}, 023 (2014).
	
	
	\bibitem{Dvali:2012en} 
	G.~Dvali and C.~Gomez,
	Eur.\ Phys.\ J.\ C {\bf 74}, 2752 (2014).
	
	\bibitem{Dvali:2017nis} 
	G.~Dvali,
	Phys.\ Rev.\ D {\bf 97}, no. 10, 105005 (2018).
	
	\bibitem{Dvali:2018xpy} 
	G.~Dvali,
	arXiv:1810.02336 [hep-th].	
	
	
	
	\bibitem{Dvali:2013vxa} 
	G.~Dvali, D.~Flassig, C.~Gomez, A.~Pritzel and N.~Wintergerst,
	Phys.\ Rev.\ D {\bf 88},  124041 (2013).
	
	\bibitem{Hayden:2007cs} 
	P.~Hayden and J.~Preskill,
	JHEP {\bf 0709}, 120 (2007).
	
	
	\bibitem{Sekino:2008he} 
	Y.~Sekino and L.~Susskind,
	JHEP {\bf 0810}, 065 (2008).
	
	\bibitem{Casadio:2014vja} 
	R.~Casadio, A.~Giugno, O.~Micu and A.~Orlandi,
	Phys.\ Rev.\ D {\bf 90},  084040 (2014).
	
	\bibitem{Casadio:2015bna} 
	R.~Casadio, A.~Giugno and A.~Orlandi,
	Phys.\ Rev.\ D {\bf 91}, no. 12, 124069 (2015).
	
	\bibitem{Casadio:2017cdv} 
	R.~Casadio, A.~Giugno, A.~Giusti and M.~Lenzi,
	Phys.\ Rev.\ D {\bf 96}, no. 4, 044010 (2017).
	
	\bibitem{Casadio:2018qeh} 
	R.~Casadio, M.~Lenzi and O.~Micu,
	Phys.\ Rev.\ D {\bf 98}, no. 10, 104016 (2018).
	
	\bibitem{Casadio:2019cux} 
	R.~Casadio, M.~Lenzi and O.~Micu,
	Eur.\ Phys.\ J.\ C {\bf 79}, no. 11, 894 (2019).
	
	\bibitem{Buoninfante:2019fwr} 
	L.~Buoninfante, G.~G.~Luciano and L.~Petruzziello,
	Eur.\ Phys.\ J.\ C {\bf 79}, no. 8, 663 (2019).
	
	
	\bibitem{Giusti:2019wdx} 
	A.~Giusti,
	Int.\ J.\ Geom.\ Meth.\ Mod.\ Phys.\  {\bf 16}, no. 03, 1930001 (2019).
	
	\bibitem{Mathur:2005zp} 
	S.~D.~Mathur,
	Fortsch.\ Phys.\  {\bf 53}, 793 (2005).
	
	\bibitem{cardoso-2016b} V.~Cardoso, E.~Franzin and P.~Pani,
	Phys.\ Rev.\ Lett.\  {\bf 116},  171101 (2016)
	Erratum: [Phys.\ Rev.\ Lett.\  {\bf 117}, 089902 (2016)].
	
	\bibitem{Cardoso:2016oxy} 
	V.~Cardoso, S.~Hopper, C.~F.~B.~Macedo, C.~Palenzuela and P.~Pani,
	Phys.\ Rev.\ D {\bf 94},  084031 (2016).
	
	
	\bibitem{Abedi:2016hgu}
	J.~Abedi, H.~Dykaar and N.~Afshordi,
	Phys.\ Rev.\ D {\bf 96},  082004 (2017).
	
	
	\bibitem{Dvali:2010bf} 
	G.~Dvali and C.~Gomez,
	arXiv:1005.3497 [hep-th].
	
	\bibitem{Dvali:2010jz} 
	G.~Dvali, G.~F.~Giudice, C.~Gomez and A.~Kehagias,
	JHEP {\bf 1108}, 108 (2011).
	
	
	\bibitem{Dvali:2016ovn} 
	G.~Dvali,
	Subnucl.\ Ser.\  {\bf 53}, 189 (2017).
	
	\bibitem{Dvali:2011th} 
	G.~Dvali, C.~Gomez and A.~Kehagias,
	JHEP {\bf 1111}, 070 (2011).
	
	\bibitem{Dvali:2014ila} 
	G.~Dvali, C.~Gomez, R.~S.~Isermann, D.~Lüst and S.~Stieberger,
	Nucl.\ Phys.\ B {\bf 893}, 187 (2015).
	
	
	
	\bibitem{Thorne:1972ji} 
	K.~S.~Thorne,
	``Nonspherical Gravitational Collapse: A Short Review,''
	in "J R Klauder, Magic Without Magic", San Francisco 1972, 231-258.
	
	
	\bibitem{Casadio:2016zpl} 
	R.~Casadio, A.~Giugno and A.~Giusti,
	Phys.\ Lett.\ B {\bf 763}, 337 (2016).	
	
	
	\bibitem{Bogolyubov:1947zz} 
	N.~N.~Bogolyubov,
	J.\ Phys.\ (USSR) {\bf 11}, 23 (1947)
	[Izv.\ Akad.\ Nauk Ser.\ Fiz.\  {\bf 11}, 77 (1947)].
	
	
	
	\bibitem{Dvali:2015aja} 
	G.~Dvali,
	Fortsch.\ Phys.\  {\bf 64}, 106 (2016).
	
	%
	%
	%
	
	\bibitem{Casadio:2013tma} 
	R.~Casadio,
	arXiv:1305.3195 [gr-qc].
	
	
	\bibitem{Casadio:2016fev} 
	R.~Casadio, A.~Giugno and A.~Giusti,
	Gen.\ Rel.\ Grav.\  {\bf 49}, no. 2, 32 (2017).	
	
	
	\bibitem{tHooft:1984kcu} 
	G.~'t Hooft,
	Nucl.\ Phys.\ B {\bf 256}, 727 (1985).
	
	
	
	\bibitem{Susskind:1993if} 
	L.~Susskind, L.~Thorlacius and J.~Uglum,
	Phys.\ Rev.\ D {\bf 48}, 3743 (1993).
	
	\bibitem{Casadio:2020ueb} 
	R.~Casadio, M.~Lenzi and A.~Ciarfella,
	arXiv:2002.00221 [gr-qc].
	
	\bibitem{Regge:1957td}
	T.~Regge and J.~A.~Wheeler,
	Phys. Rev. \textbf{108} (1957), 1063-1069.
	
	
	\bibitem{Zerilli:1970se}
	F.~J.~Zerilli,
	Phys. Rev. Lett. \textbf{24} (1970), 737-738.
	
	
	
	\bibitem{Carballo-Rubio:2018jzw} 
	R.~Carballo-Rubio, F.~Di Filippo, S.~Liberati and M.~Visser,
	Phys.\ Rev.\ D {\bf 98}, no. 12, 124009 (2018).
	
	\bibitem{Guo:2017jmi} 
	B.~Guo, S.~Hampton and S.~D.~Mathur,
	JHEP {\bf 1807}, 162 (2018).
	
	\bibitem{Cardoso:2017njb}
	V.~Cardoso and P.~Pani,
	[arXiv:1707.03021 [gr-qc]].
	
	
	\bibitem{Oshita:2019sat}
	N.~Oshita, Q.~Wang and N.~Afshordi,
	JCAP \textbf{04}, 016 (2020).
	
	\bibitem{Cardoso:2017cqb} 
	V.~Cardoso and P.~Pani,
	Nat.\ Astron.\  {\bf 1}, 586 (2017).
	
	\bibitem{Holdom:2016nek}
	B.~Holdom and J.~Ren,
	Phys. Rev. D \textbf{95}, no.8, 084034 (2017).
	
	
	\bibitem{Abedi:2018npz} 
	J.~Abedi and N.~Afshordi,
	JCAP {\bf 1911},  010 (2019).
	
	\bibitem{Buoninfante:2019swn} 
	L.~Buoninfante and A.~Mazumdar,
	Phys.\ Rev.\ D {\bf 100}, no. 2, 024031 (2019).
	
	\bibitem{Buoninfante:2019teo} 
	L.~Buoninfante, A.~Mazumdar and J.~Peng,
	Phys.\ Rev.\ D {\bf 100}, no. 10, 104059 (2019).
	
	\bibitem{Cardoso:2019rvt}
	V.~Cardoso and P.~Pani,
	Living Rev.\ Rel.\  {\bf 22},  4 (2019).
	
	\bibitem{Foit:2016uxn} 
	V.~F.~Foit and M.~Kleban,
	Class.\ Quant.\ Grav.\  {\bf 36}, no. 3, 035006 (2019).
	
	\bibitem{Maggio:2017ivp}
	E.~Maggio, P.~Pani and V.~Ferrari,
	Phys. Rev. D \textbf{96}, no.10, 104047 (2017).
	
	\bibitem{Maggio:2018ivz}
	E.~Maggio, V.~Cardoso, S.~R.~Dolan and P.~Pani,
	Phys. Rev. D \textbf{99}, no.6, 064007 (2019).
	
	\bibitem{Testa:2018bzd}
	A.~Testa and P.~Pani,
	Phys. Rev. D \textbf{98}, no.4, 044018 (2018).
	
	\bibitem{Cardoso:2019apo} 
	V.~Cardoso, V.~F.~Foit and M.~Kleban,
	JCAP {\bf 1908}, 006 (2019).
	
	\bibitem{Coates:2019bun}
	A.~Coates, S.~H.~Völkel and K.~D.~Kokkotas,
	Phys. Rev. Lett. \textbf{123} (2019) no.17, 171104.
	
	
	\bibitem{Wang:2019rcf} 
	Q.~Wang, N.~Oshita and N.~Afshordi,
	Phys.\ Rev.\ D {\bf 101}, no. 2, 024031 (2020).
	
	
	\bibitem{Maggio:2019zyv}
	E.~Maggio, A.~Testa, S.~Bhagwat and P.~Pani,
	Phys. Rev. D \textbf{100}, no.6, 064056 (2019).

\bibitem{Buoninfante:2020cqz}
L.~Buoninfante, G.~Lambiase, G.~G.~Luciano and L.~Petruzziello,
Eur. Phys. J. C \textbf{80} (2020) no.9, 853
	
	
	\bibitem{Abedi:2020ujo} 
	J.~Abedi, N.~Afshordi, N.~Oshita and Q.~Wang,
	arXiv:2001.09553 [gr-qc].
	
	\bibitem{Salvio:2019llz}
	A.~Salvio and H.~Veermäe,
	JCAP \textbf{02}, 018 (2020).
	
	\bibitem{Konoplya:2018yrp}
	R.~A.~Konoplya, Z.~Stuchlík and A.~Zhidenko,
	Phys.\ Rev.\ D {\bf 99}, no. 2, 024007 (2019).
	
	\bibitem{Bronnikov:2019sbx}
	K.~A.~Bronnikov and R.~A.~Konoplya,
	Phys.\ Rev.\ D {\bf 101}, no. 6, 064004 (2020).
	
	\bibitem{Wang:2018mlp}
	Y.~T.~Wang, Z.~P.~Li, J.~Zhang, S.~Y.~Zhou and Y.~S.~Piao,
	Eur. Phys. J. C \textbf{78}, no.6, 482 (2018).
	
	\bibitem{Wang:2019szm}
	Y.~T.~Wang, J.~Zhang, S.~Y.~Zhou and Y.~S.~Piao,
	Eur. Phys. J. C \textbf{79}, no.9, 726 (2019).
	
	\bibitem{Maggio:2020jml}
	E.~Maggio, L.~Buoninfante, A.~Mazumdar and P.~Pani,
	Phys. Rev. D \textbf{102} (2020) no.6, 064053.
	
	
	\bibitem{Saraswat:2019npa}
	K.~Saraswat and N.~Afshordi,
	JHEP \textbf{04}, 136 (2020).
	
	
	\bibitem{Conklin:2017lwb} 
	R.~S.~Conklin, B.~Holdom and J.~Ren,
	Phys.\ Rev.\ D {\bf 98},  044021 (2018).
	
	\bibitem{Ashton:2016xff}
	G.~Ashton {\it et al.},
	arXiv:1612.05625 [gr-qc]; J.~Abedi, H.~Dykaar and N.~Afshordi,
	arXiv:1701.03485 [gr-qc].
	
	\bibitem{Westerweck:2017hus} 
	J.~Westerweck {\it et al.},
	Phys.\ Rev.\ D {\bf 97},  124037 (2018);  J.~Abedi, H.~Dykaar and N.~Afshordi,
	arXiv:1803.08565 [gr-qc].
	
	
	\bibitem{tHooft:1973alw} 
	G.~'t Hooft,
	Nucl.\ Phys.\ B {\bf 72}, 461 (1974).
	
	\bibitem{Dvali:2020wqi}
	G.~Dvali,
	arXiv:2003.05546 [hep-th].
	
	\bibitem{Burgess:2018pmm} 
	C.~P.~Burgess, R.~Plestid and M.~Rummel,
	JHEP {\bf 1809}, 113 (2018).
	
	\bibitem{Torres:2018dso}
	T.~Torres, S.~Patrick, M.~Richartz and S.~Weinfurtner,
	Phys. Rev. Lett. \textbf{125} (2020) no.1, 011301.

	
\end{thebibliography}
\end{document}